\documentclass[prd,aps,floatfix,nofootinbib,twocolumn,tightenlines,showpacs]{revtex4}
\usepackage{dcolumn}
\usepackage{latexsym}
\usepackage{graphicx}
\usepackage{bm}

\newcommand{\be}{\begin{equation}}
\newcommand{\ee}{\end{equation}}
\newcommand{\bea}{\begin{eqnarray}}
\newcommand{\eea}{\end{eqnarray}}

\def\csw{c_{\text{SW}}}
\def\Eq#1{Eq.~(\ref{#1})}
\def\wo{\widehat O}

\begin{document}
\title{
Improvement via hypercubic smearing in triplet and sextet QCD}
\author{Yigal Shamir}
\author{Benjamin Svetitsky}
\author{Evgeny Yurkovsky}
\affiliation{Raymond and Beverly Sackler School of Physics and Astronomy, Tel~Aviv University, 69978 Tel~Aviv, Israel}

\begin{abstract}
We study non-perturbative improvement in SU(3) lattice gauge theory coupled to
fermions in the fundamental and two-index symmetric representations.
Our lattice action is defined with hypercubic smeared links incorporated into the Wilson--clover fermion kernel.
Using standard Schr\"odinger-functional techniques we estimate the clover coefficient $\csw$
and find that discretization errors are much smaller than in thin-link theories.
\end{abstract}

\maketitle

\section{Introduction}

The improvement of a lattice action is meant to reduce the effects of lattice artifacts and thus to bring calculated quantities closer to their continuum limits.
In our work on the SU(3) gauge theory with sextet fermions \cite{Shamir:2008pb, DeGrand:2008kx,DeGrand:2010na,Svetitsky:2010zd} we have adopted normalized hypercubic (nHYP) smearing \cite{Hasenfratz:2001hp,Hasenfratz:2007rf} in the expectation that it would yield reliable results on fairly coarse lattices.
We found, indeed \cite{DeGrand:2010na}, that the smeared-link action
keeps the critical hopping parameter $\kappa_c(\beta)$ of the Wilson--clover
action much closer to its continuum value of $1/8$ even for strong bare couplings $g_0^2\equiv6/\beta$.
It also
pushes a first-order phase transition towards stronger bare couplings, thus allowing calculation of the running coupling $g^2_{\text{SF}}$
in a regime where both $g_0^2$ and $g^2_{\text{SF}}$ are quite strong.
Moreover, the smeared plaquette averages are much closer to unity than the ``thin-link'' plaquettes, a clear sign that the gauge field is being effectively
smoothed.

In this note we show that nHYP improvement extends to the axial Ward identity (AWI).
In a calculation involving Wilson fermions, the AWI is frequently used to determine $\kappa_c$ by demanding that the quark mass $m$ be zero, a consequence of conservation of the isovector axial current.
One can adjust the improvement coefficient
$\csw$, which multiplies the clover term in the fermion action \cite{Sheikholeslami:1985ij}, in order to minimize errors in the AWI itself.
One sign of such errors is the sensitivity of $m$ to the location where the AWI is measured.
L\"uscher and collaborators \cite{Luscher:1996sc,Luscher:1996ug} proposed a procedure of measuring $m$ at points of the lattice that are inequivalent because of the boundary conditions used in the Schr\"odinger Functional (SF) method.
Demanding that $m=0$ at two such points gives conditions for fixing $\kappa_c$ and $\csw$.

L\"uscher {\em et al.}~\cite{Luscher:1996ug} applied the AWI criterion to the quenched SU(3) gauge theory, defining the axial current using thin links.
While $\csw=1$ at tree level, they found numerically
that $\csw\simeq1.8$ is required for $\beta=6$.
Jansen and Sommer \cite{Jansen:1998mx} did a similar calculation for QCD with thin-link Wilson fermions ($N_f=2$) and again found that a large value of $\csw-1$ is required.
Defining the current with nHYP-smeared links,
Hoffmann, Hasenfratz, and Schaefer~\cite{Hoffmann:2007nm} reduced the
required $\csw$ in the quenched theory from 1.55 to 1.05 at $\beta\simeq6.4$.
This is dramatic evidence for the claim that nHYP smearing brings the theory much closer to the continuum limit.

In this paper we present a calculation in theories with $N_f=2$ dynamical triplet
and sextet quarks.
For quarks in the triplet representation, we compare the thin-link and smeared-link theories and find results as dramatic as in the quenched theory.
We find small violations of the AWI in the smeared-link sextet theory as well.
We conclude that setting $\csw$ to 1 is adequate when nHYP smearing is used.

The procedure of keeping $\csw=1$ while using smeared links was advocated for triplet QCD in Ref.~\cite{Capitani:2006ni} and subsequently tested in large-scale calculations~\cite{Durr:2008rw, Kurth:2010yk}.  For a determination of $\csw$ with partial stout smearing, see~\cite{Cundy:2009yy}.

\section{Determination of $\csw$} \label{sec:Csw}

We follow closely the method of Refs.~\cite{Luscher:1996sc,Luscher:1996ug,Jansen:1998mx,Hoffmann:2007nm}.
The $O(a)$-improved action is \cite{Sheikholeslami:1985ij}
\be
	S = S_{\text W}[U,\bar\psi,\psi] +
	a^5\csw\sum_x\bar\psi\frac i4\sigma_{\mu\nu}F_{\mu\nu}\psi,
	\label{eq:action}
\ee
where $S_{\text W}$ is the conventional Wilson action comprised of the plaquette gauge action and the fermion hopping term.
The second term in \Eq{eq:action} is the clover term, wherein $F_{\mu\nu}$
is the lattice-discretized field strength.
The gauge links in the hopping and clover terms are ``fat links,'' defined via nHYP smearing as described in Ref.~\cite{Hasenfratz:2001hp} and with the same smearing parameters.

On a lattice with $L^3\times T$ sites, we impose Dirichlet boundary conditions on the gauge fields on the temporal boundaries of the lattice,
\be
U_k(x)\big|_{x_0=0}=\exp C_k,\qquad
U_k(x)\big|_{x_0=T}=\exp C'_k,
\ee
with the asymmetric choice%
\footnote{Note the appearance of $T$ in \Eq{bkgd}.
The authors of Refs.~\cite{Luscher:1996sc,Luscher:1996ug,Jansen:1998mx,Hoffmann:2007nm} use $L$ instead, which means that for the lattice used here our background field is weaker than theirs.
}
\be
C_k=\frac i{6T}\text{diag}(-\pi,0,\pi),\qquad
C'_k=\frac i{6T}\text{diag}(-5\pi,2\pi,3\pi).
\label{bkgd}
\ee
The fermion boundary conditions are homogeneous,
\begin{eqnarray}
  P_- \psi(t=0) &= P_+ \psi(t=T) &= 0,
\nonumber\\[2pt]
 \bar\psi(t=0) P_+ &= \bar\psi(t=T) P_- &= 0,
\label{SFrho}
\end{eqnarray}
with $P_\pm=\frac12(1\pm\gamma_0)$.
The gauge-invariant boundary fields that
can be used as wall sources are
\begin{eqnarray}
\zeta &=&\sum_{\bf x} U_0({\bf x},t=0) P_+ \psi({\bf x},t=1)\\
\bar\zeta&=& \sum_{\bf x} \bar\psi({{\bf x},t=1}) P_- U^\dagger_0({\bf x},t=0)\\
\zeta'&=&\sum_{\bf x} U^\dagger_0({\bf x},t=T-1) P_- \psi({\bf x},t=T-1)\\
\bar\zeta'&=& \sum_{\bf x} \bar\psi({{\bf x},t=T-1}) P_+ U_0({\bf x},t=T-1).
\label{SFzeta}
\end{eqnarray}
(These are the same as those used by \cite{Luscher:1996sc}, but written in
explicit form.)
The gauge fields are periodic in the spatial directions, while the fermion fields satisfy $\psi(L)=e^{i\pi/5}\psi(0)$ \cite{Luscher:1996sc}.

The axial Ward identity
\be
\partial_\mu A_{\text{imp}}^{\mu a} = 2mP^a
\label{eq:PCAC}
\ee
gives a definition of the quark mass $m$.
Equation~(\ref{eq:PCAC}) contains the $O(a)$-improved axial current,
\be
A_{\text{imp}}^{\mu a}=A^{\mu a}
+c_Aa\partial^\mu P^a.
\ee
The pseudoscalar density and the unimproved axial current are defined by the local products
\be
P^a=\bar\psi\gamma_5\frac{\tau^a}2\psi,\qquad
A_{\mu}^{a}=\bar\psi\gamma_5\gamma_\mu\frac{\tau^a}2\psi.
\ee
In practice, one evaluates correlation functions of \Eq{eq:PCAC},
\be
	\partial_{\mu}\langle A_{\text{imp}}^{\mu a}(x)\wo\rangle =
	2m\langle P^a(x)\widehat{O}\rangle,
	\label{eq:pcac}
\ee
where $\wo$ is any operator located at non-zero distance from $x$.
It is convenient \cite{Luscher:1996sc} to use the pseudoscalar field made of the boundary operators to define wall sources $\wo$ and $\wo'$ at $t=0$ and $t=T$, respectively, {\em viz.,}
\be
\wo^a = \bar\zeta \gamma_5 \frac{\tau^a}2 \zeta,\qquad \wo^{\prime a} = \bar\zeta' \gamma_5 \frac{\tau^a}2 \zeta'.
\ee
Thus we define the correlation functions
\begin{eqnarray}
f_P(x_0)&=&- \frac13\langle P^a(x_0)\widehat{O}^a\rangle\\
f_A(x_0)&=&- \frac13\langle A_0^a(x_0)\widehat{O}^a\rangle,
\end{eqnarray}
which depend only on $x_0$ by translation invariance of the wall sources.
The spatial derivatives in \Eq{eq:pcac} vanish similarly,
whence one obtains an estimate for $m$,
\be
	m(x_0) = \frac{\partial_{0}f_A(x_0)+
	c_A a\partial_{0}^2f_P(x_0)}{2f_P(x_0)},
	\label{eq:mass}
\ee
where for $\partial_0$ we use a symmetric derivative and $\partial_{0}^2$
is the nearest-neighbor second derivative.
Using $\wo'$ we define analogously
\begin{eqnarray}
f_P'(T-x_0)&=&- \frac13\langle P^a(x_0)\wo^{\prime a}\rangle\\
f_A'(T-x_0)&=&- \frac13\langle A_0^a(x_0)\wo^{\prime a}\rangle,
\end{eqnarray}
which leads to an alternative estimate $m'(x_0)$, defined by the parallel of \Eq{eq:mass} in terms of $f_P',f_A'$.
If the AWI is respected by the improved action then one would expect that $m$ and $m'$ are independent of $x_0$ and equal to each other.

Equation~(\ref{eq:mass}) and its primed counterpart still contain the unknown coefficient $c_A$.
An alternative definition of the mass eliminates this dependence.
If we write \Eq{eq:mass} as
\be
m(x_0)=r(x_0)+c_As(x_0),
\ee
and similarly for $m',r',s'$,
the alternative is
\be
	M(x_0,y_0) = r(x_0)-s(x_0)\frac{r(y_0)-r'(y_0)}{s(y_0)-s'(y_0)},
	\label{eq:new_m}
\ee
which differs from $m,m'$ in $O(a^2)$.
We also define the quantity $M'$ by exchanging primed and unprimed variables in \Eq{eq:new_m}.
Then, still following Refs.~\cite{Luscher:1996ug,Jansen:1998mx,Hoffmann:2007nm},
we define the measure of residual violation of the AWI to be
\be
	\Delta M = M\left(\frac34T,\frac14T\right)-M'\left(\frac34T,\frac14T\right).
	\label{eq:dm}
\ee

\section{Numerical results}

\begin{figure}[ht]
\includegraphics[width=\columnwidth,clip]{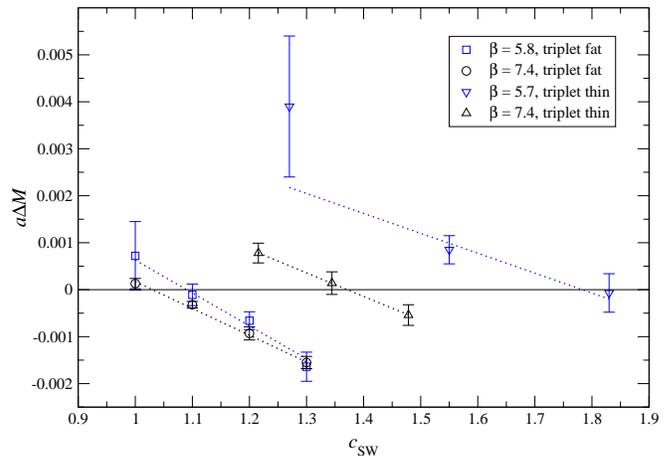}
\caption{Determination of $\csw$ for the fat- and thin-link theories with triplet quarks, using the $\Delta M$ data in Tables~\ref{tab:dm_csw_fund_table} and~\ref{tab:dm_csw_fund_thin_table}.
The curves are linear fits whose slopes and intercepts are given in Table~\ref{tab:optimal_slope}.
\label{fig:dm_csw_all_fit}}
\end{figure}

\begin{figure}[ht]\includegraphics[width=\columnwidth,clip]{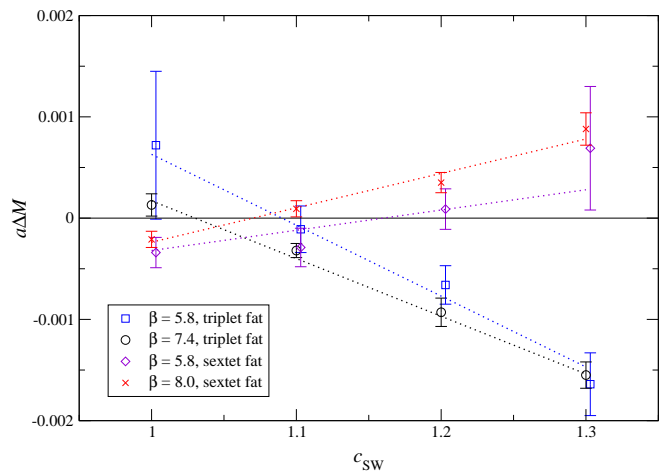}
\caption{Data for the fat-link theory with sextet quarks (Table~\ref{tab:dm_csw_table}), plotted together with
the fat-link triplet data from Fig.~\ref{fig:dm_csw_all_fit}.
The linear fits shown have the
parameters listed in Table~\ref{tab:optimal_slope}.
Data points for $\beta=5.8$ have been given slight horizontal shifts for clarity.
\label{fig:fits}}
\end{figure}

All our calculations were performed on lattices with $L=8$, $T=16$.
For triplet as well as for sextet quarks we chose two values of $\beta$,
in the weak and intermediate coupling regions.
Varying $\csw$, we calculated $\Delta M$ as defined in \Eq{eq:dm}.
Our results are displayed in Tables~\ref{tab:dm_csw_fund_table}--\ref{tab:optimal_slope} and in Figs.~\ref{fig:dm_csw_all_fit} and~\ref{fig:fits}.

\begin{table}[htb]
\begin{ruledtabular}
\begin{tabular}{dddd}
\beta & \kappa & \csw & a\Delta M \times 10^4 \\
\hline
7.4 & 0.1255 & 1   &   1. 3(11) \\
    &        & 1.1 &  -3.2(7)  \\
    &        & 1.2 &  -9.3(14) \\
    &        & 1.3 & -15.5(13) \\[3pt]
5.8 & 0.1267& 1   &   7.2(73)  \\
    &        & 1.1 &  -1.1(23) \\
    &        & 1.2 &  -6.6(19) \\
    &        & 1.3 & -16.4(31) \\
\end{tabular}
\end{ruledtabular}
\caption{Results of $\Delta M$ calculations on $8^3\times16$ lattice. SU(3) gauge theory with $N_f = 2$ fermions in the fundamental representation, fat links.
2000 trajectories were run at $\beta=7.4$ for each value of $\csw$, and 3000
at $\beta=5.8$.
\label{tab:dm_csw_fund_table}}
\end{table}
\begin{table}[htb]
\begin{ruledtabular}
\begin{tabular}{dddd}
\beta & \kappa & \csw & a\Delta M \times 10^4 \\
\hline
7.4 & 0.1346  & 1.2156 &   7.8(21) \\
    & 0.1334  & 1.3445 &   1.4(24) \\
    & 0.13245 & 1.4785 &  -5.4(22) \\[3pt]
5.7 & 0.14133 & 1.27   &  39.(15)  \\
    & 0.13786 & 1.55   &  8.5(30)  \\
    & 0.13433 & 1.83   &  -0.7(41) \\
\end{tabular}
\end{ruledtabular}
\caption{$\Delta M$ for thin-link fermions in the fundamental representation, for comparison with Table~\ref{tab:dm_csw_fund_table}.
Values of $\beta$, $\kappa$, and $\csw$ were chosen as in Ref.~\cite{Jansen:1998mx}.
4000 trajectories were run at $\beta=7.4$ for each value of $\csw$, and 18000
at $\beta=5.7$.
\label{tab:dm_csw_fund_thin_table}}
\end{table}
\begin{table}[htb]
\begin{ruledtabular}
\begin{tabular}{dddd}
\beta & \kappa & \csw & a\Delta M \times 10^4 \\
\hline
8   & 0.12688 & 1   & -2.1(8)  \\
    &         & 1.1 &  0.9(8)  \\
    &         & 1.2 &  3.5(10) \\
    &         & 1.3 &  8.8(16) \\[3pt]
5.8 & 0.12835 & 1   & -3.4(15) \\
    &         & 1.1 & -2.9(19) \\
    &         & 1.2 &  0.9(20) \\
    &         & 1.3 &  6.9(61) \\
\end{tabular}
\caption{As in Table \ref{tab:dm_csw_fund_table}, but for fat-link fermions in the sextet representation. 4000 trajectories were run for each $\beta,\csw$.
\label{tab:dm_csw_table}}
\end{ruledtabular}
\end{table}


\begin{table}[htb]
\begin{ruledtabular}
\begin{tabular}{lldd}
  Theory & smearing  &  \csw^*    &  \text{Slope}      \\
	\hline
$\beta=7.4$, triplet & thin links & 1.37(3)  & -0.005(1)            \\
$\beta=7.4$, triplet & fat links  & 1.04(1)  & -0.0057(5)            \\[3pt]
$\beta=5.7$, triplet & thin links & 1.78(8)  & -0.004(2)             \\
$\beta=5.8$, triplet & fat links  & 1.10(3)  & -0.007(2)             \\[3pt]
$\beta=8$,    sextet & fat links  & 1.07(2)  &  0.0033(5)            \\
$\beta=5.8$,  sextet & fat links  & 1.17(6) &  0.002(1)
\end{tabular}
\end{ruledtabular}
\caption{Results of fitting $\Delta M(\csw)$ to a straight line for each theory: the value of $\csw$ where $\Delta M=0$, and the fitted slope.}
\label{tab:optimal_slope}
\end{table}

For the fat-link theories, we fixed $\kappa$ for each $\beta$ by demanding that $r(T/2)=0$
at $\csw=1$, that is,
by setting to zero the unimproved quark mass.
This is an alternative to requiring, say, $M(x_0,y_0)=0$ for some $x_0,y_0$;
it follows on the observation that the second term in \Eq{eq:new_m} is generally
small and recognizes the fact that the
AWI (\ref{eq:PCAC}) should hold for nonzero mass as well.
In comparing to other work we note that the $\kappa$-dependence of $\Delta M$
has been seen to be very weak~\cite{Luscher:1996ug,Jansen:1998mx}.
For the same reason, we did not vary $\kappa$ for the fat-link theories as we varied $\csw$ at given $\beta$.

The optimal value $\csw^*(\beta)$ is determined by demanding $\Delta M=0$.
We do this via linear fits to $\Delta M(\csw)$, with the results shown in Table~\ref{tab:optimal_slope} and plotted in the figures.

In the theory with triplet fermions we compare the fat-link results to $\Delta M$ in the thin-link theory, calculated at values of $\beta$, $\kappa$, and $\csw$ used by Jansen and Sommer~\cite{Jansen:1998mx}.  [In this case, $\kappa$ was shifted with $\csw$ to keep $M(\frac12T,\frac14T)=0$.]
It is clear that for $\csw\simeq1$ the fat-link action gives much smaller values of $\Delta M$ for both values of $\beta$.
The fat-link action also gives $\csw^*(\beta)$ very close to unity, even at the stronger coupling $\beta=5.8$, for both the triplet and the sextet theory.

\section{Discussion}

Figs.~\ref{fig:dm_csw_all_fit} and~\ref{fig:fits} and Table~\ref{tab:optimal_slope} show that
discretization errors in both fat-link theories are generally
small, as reflected in the value of $\Delta M$ at $\csw=0$.
In comparing thin- and fat-link fermions for the triplet theory, we find that the slope of $\Delta M$ vs.~$\csw$ is similar but that
the nonperturbatively determined coefficient $\csw^*$
is much closer to one for the smeared links.

Our comparison between thin- and fat-link theories
uses results obtained
at (roughly) the same values of the bare coupling.
One could compare instead at points of similar physics by demanding that
the values of $\beta$ give, for example, the same string tension.
This would constitute a comparison of the different
discretizations at equal lattice spacing.
Our results, however, make this elaborate exercise unnecessary.
In contrast with the triplet thin-link results,
the fat-link results change very little in going from $\beta=7.4$ to~5.8.
Thus, had we tuned our bare couplings
to the slightly different values needed to reproduce the
lattice spacings of Ref.~\cite{Jansen:1998mx},
there would be no qualitative change in the conclusions.

Turning to the sextet theory, one might be concerned by the small slope of
$\Delta M(\csw)$, which acts to enlarge the error bar on $\csw^*$.
Is this due to having chosen a quantity with low sensitivity to $\csw$?
Let us consider the tree-level value of $\Delta M$, denoted $\Delta M^{(0)}$.
This quantity sets a scale for AWI violation by the
background field on the finite lattice.
It is easily determined numerically, by calculating the Green functions in
Sec.~\ref{sec:Csw} in the single gauge configuration that is the classical
solution to the boundary conditions, namely, a constant electric field.
(We set $\csw=1$ and $\kappa=1/8$.)
In the triplet theory we obtain $a\Delta M^{(0)}=0.00025$,
whereas for the sextet theory, $a\Delta M^{(0)}=-0.00065$.
In the sextet theory the fluctuations of the dynamical fields
{\em weaken\/} $\Delta M$ at $\csw=1$ considerably from its
tree-level value.  The observation that $\Delta M(\csw=1)$
is small on the scale set by $\Delta M^{(0)}$
means that there is no reason to vary $\csw$ away from 1.

The authors of Refs.~\cite{Luscher:1996ug,Jansen:1998mx,Hoffmann:2007nm}
fixed $\csw^*$ by demanding $\Delta M=\Delta M^{(0)}$.
This is an attempt to determine $\csw^*$ in infinite volume, by supposing
that the finite-volume corrections to the dynamical $\Delta M$ and to
the tree-level $\Delta M^{(0)}$ are the same.
Since this assumption is not yet supported by any study of volume dependence,
we eschew this procedure in favor of demanding $\Delta M=0$.
In other words, we determine $\csw^*$ for our volume without
making any statement about the $L\to\infty$ limit.

The smallness of discretization effects we report for the fat-link theories
is consistent with the reduced discretization errors found in
other observables in the SU(3)/sextet theory~\cite{DeGrand:2010na} as well as
in the SU(2)/adjoint theory~\cite{SU2}.%
\footnote{For a similar determination of $\csw$ in the SU(2) gauge theory with thin-link
fermions in the fundamental and adjoint representations, see \cite{Mykkanen:2010ym,Karavirta:2011mv}.}
This lends support to our decision to stick with the tree-level value,
$\csw=1$.  Indeed, thanks to asymptotic freedom,
the continuum limit of a lattice gauge theory
with a given (Dirac-) fermion content is completely determined once
the fermion masses are fixed.  Improvement is the art of reducing
discretization errors when the lattice spacing is nonzero.
Since it does not change the continuum limit, improvement is always
at one's discretion, never mandatory.  Our experience is that,
thanks to the fat-link action, using $\csw=1$ provides as much
improvement as we need.

\begin{acknowledgments}
We thank Stefan Schaefer for correspondence and Thomas DeGrand for much discussion.
B.~S. and Y.~S. thank the University of Colorado for hospitality.
This work was supported in part by the Israel Science Foundation
under grant no.~423/09.
\end{acknowledgments}

\end{document}